\documentstyle[epsf]{l-aa} 

\begin{document}

\newcommand{\Kkms}{\,{\rm K\,km/s} }
\newcommand{\WHI}{\,{W_{\rm HI}} } 

\newcommand{\mic}{\,{\rm \mu m} } 
\newcommand{\pac}{\,{\rm pc} } 

\newcommand{\Wmsr}{\,{\rm W/m^2/sr} }
\newcommand{\Wcmsr}{\,{\rm W/cm^2/sr} }
\newcommand{\Wcm}{\,{\rm W/cm^2} }
\newcommand{\cmun}{\,{\rm cm^{-1}} }
\newcommand{\Mj}{\,{\rm MJy/sr} }

\thesaurus{Sect. 8: Diffuse matter in space ,
            09.03.1;        % ISM:clouds 
            09.04.1;        % ISM:dust,extinction
            13.09.3;       % Infrared:ISM:continuum 
            09.19.1;        % ISM:structure
            13.18.3;	  % Radio-continuum:ISM
            09.07.1 }     % ISM:general	

\title{The interstellar cold dust observed by COBE} 
\subtitle{} 

   \author{G. Lagache  
          \inst{1} \and 
          A. Abergel 
          \inst{1} \and 
          F. Boulanger 
          \inst{1} \and 
           J.-L. Puget 
          \inst{1}}
 
 \offprints{G. Lagache } 
 
\institute{$^1$ Institut d'Astrophysique Spatiale, B\^at.  121, 
Universit\'e Paris XI, F--91405 Orsay Cedex, France}

\date{Received 4 June 1997; accepted 11 December 1997} 
%\date{latest revision 18-12-97} 
 
   \maketitle
 
   \begin{abstract}
Using DIRBE and FIRAS maps at high latitude ($|b|>10\degr$) we derive the
spatial distribution of the dust temperature
 associated with the diffuse cirrus and the dense molecular clouds. 
For a $\nu^2$ emissivity law, we find that the equilibrium dust temperature of the cirrus
is about 17.5 K with only small variations over the high latitude sky.
Comparison of the far
Infrared DIRBE maps shows the presence of a colder emission component with a temperature 
around 15 K, assuming a $\nu^2$ emissivity law.
The lowest values of the temperature found in the cold regions ($\sim13$ K) are 
compatible with the results recently obtained for dense cores in star forming regions by the 
balloon-borne experiment SPM-PRONAOS (Ristorcelli
et al., 1996, 1998, Serra et al., 1997).
This cold component is in particular present in the direction of known 
molecular complexes with low star forming activity such as Taurus. 
The association between the cold component and molecular clouds is further demonstrated
by the fact that all sky pixels with significant cold emission have an 
excess IR emission with respect to the high latitude IR/HI correlation.
We have deduced a threshold value of the column density, 
N$_{HI}$=2.5 $10^{20}$ H cm$^{-2}$, below which cold dust 
is not detected within the FIRAS beam of $\sim7\degr$.
We have re-examined the problem of the existence of a very cold dust component ($T\sim7$ K)
by combining DIRBE maps of the cold 
emission with FIRAS spectra, corrected for the isotropic component found in Puget et al. (1996).
 The warm and cold component deduced from the analysis of
DIRBE maps account for the Galactic FIRAS spectra with no need 
for a very cold component ($T\sim7$ K).


   \end{abstract}

\section{Introduction}

Much has been learned about the emission of the interstellar dust from IRAS
observations. To account for the Galactic energy emitted in the near to far infrared, it
is necessary to have a broad dust size distribution from large grains
down to large molecules. 
For example, D\'esert et al. (1990) (see also Draine \& Anderson 1985, 
Puget et al. 1985, Weiland et al. 1986, Siebenmorgen \& Kr\"ugel 1992
and Dwek et al. 1997) have proposed a consistent interpretation of both
the infrared emission in diffuse HI clouds and the interstellar extinction
curve using a model with three components: PAHs (Policyclic Aromatic
Hydrocarbons), very small grains (VSGs) and large grains. 
PAHs and VSGs are small enough ($a\le10$ nm) to experience significant temperature fluctuations
after photon absorption. They emit over a wide range of temperatures and dominate
the emission for $\lambda\le60\mic$.
The large grain component is the more traditional dust component historically inferred
from optical studies. These grains are in
equilibrium with the incident radiation field with a temperature
of about 17 K in the diffuse atomic medium.
The large spatial variations of the infrared spectrum over the 
wavelenghts range $12-100\mic$ have been interpreted as changes in the abundance of small
grains (Boulanger et al. 1990, Bernard et al. 1993).\\

With the DIRBE (Diffuse InfraRed Background Experiment) and FIRAS (Far InfraRed
Absolute Spectrophotometer) instruments on board the COsmic Background Explorer 
(COBE) satellite, we have a measure of the 
whole emission spectrum of the interstellar dust from the near infrared to 
millimeter wavelengths. 
 The mean Far Infrared (FIR) and submillimeter (submm) spectrum of the whole Galaxy
 was first derived by Wright et al. (1991) using the FIRAS data. They showed that
 this spectrum can be fitted by a single temperature component with an emissivity 
 index ($\alpha$) equal to 1.65 (T=23.3 K) or by a two temperature model
 assuming $\alpha$=2 ($T_{1}$=20.4 K and $T_{2}$=4.77 K), with a statistically better agreement 
 for the two component model.
More recently, Reach et al. (1995) have followed this analysis by performing
a spectral decomposition of the FIRAS data for 146 bins over the whole sky. They have identified
a very cold Galactic component (4-7 K) correlated with the 
warm component (16-21 K). The correlation strongly suggests a Galactic origin
for the very cold component.
The presence of such a component questioned our understanding of the 
physics of the FIR Galactic emitters. Only clouds 
with a central extinction $A_{v}$ higher than a few tens could sufficiently 
attenuate the heating rate for classical
interstellar grains (graphite or silicate). Such high extinction corresponds 
to very dense molecular clouds which are not ubiquitous in the Galaxy. 
The presence of 
a very cold emission component outside high extinction regions would  thus demonstrate 
the existence of a yet unknown dust component (e.g.  needle-like or fractal grains, 
or very large particles)
and/or a  departure from a $\nu^2$ power-law emissivity due for example to a sub-mm spectral feature. 
However, the Reach et al. (1995) analysis is based on total power FIRAS 
spectra which could include
a non-Galactic emission component.  
Using HI data in addition to FIRAS spectra, Puget et al. (1996) have found, 
in the residual emission after 
the removal of the HI
correlated emission, an isotropic component which could be the 
Cosmic Far Infrared BackgRound (CFIBR) due to distant galaxies. 
At $|b|>40\degr$, 
at least one third of the emission at $500\mic$ comes from the isotropic component
which should strongly contaminate the analysis
of faint regions of Reach et al. (1995). However, this isotropic component
 cannot explain the observed
correlation between the warm and very cold optical depths.

Boulanger et al. (1996) and Dwek et al. (1997) have derived 
the emission spectrum of dust using the spatial correlation of the FIRAS data with  
Galactic templates. This spectrum is insensitive to 
any isotropic terms (as the CFIBR). 
The spectrum of Dwek et al. 
was obtained by deriving the slope of the
$I(\lambda)$ vs $I(100\mic)$ correlation diagrams at high latitude. 
Boulanger et al. have used the correlation between the FIR and HI 21cm emission to 
estimate the spectrum of dust associated with HI gas.  Both spectra are very 
well-fitted by a single modified Planck curve with a $\nu^2$ emissivity law and a
temperature of 17.5 K. No significant very cold component is detected in the diffuse medium at 
high latitude.

In this paper, we present a statistical analysis of the dust temperature in the
 nearby interstellar medium seen at $|b| >10^\circ $, using DIRBE and FIRAS data.
We show that the dust in molecular clouds is colder than in diffuse atomic gas and 
argue that the detection of very cold dust is 
due to the combined emission of the isotropic background and cold molecular clouds.
In Sect. 2, we present the data we have used.
The map of the large grain temperature is derived and analysed in Sect. 3.
The DIRBE data at $\lambda \geq 60 \mu$m allow us to build the maps of the 
excess of the FIR/60$\mic$ correlation 
(Sect. 4a). These excess maps trace the cold molecular clouds
(Sect. 4b-c) and thus correspond to the cold component of the dust emission. 
The cold maps, combined with FIRAS spectra, 
point out the significant
difference in the dust temperatures in the atomic and molecular parts of 
the ISM in the whole sky (Sect. 5.1). 
The FIRAS residual emission is discussed in Sect. 5.2. 
A general discussion is presented in Sect. 6.  

\section{\label{data} Data presentation and preparation}

DIRBE is a photometer with ten bands covering the range
from $1.25$ to $240 \mic$ with 40 arcmin resolution (Silverberg et al. 1993). 
We restrict our analysis to $\lambda\ge60\mic$ because our study is focused on the 
large grain emission of the ISM. We use annual 
averaged maps because
they have a higher signal to noise ratio than maps
interpolated at the solar elongation of $90\degr$  
(see the DIRBE explanatory supplement).

The FIRAS instrument is a polarising Michelson 
interferometer with $7\degr$ resolution and two separate bands which have a
fixed spectral resolution of 0.57 cm$^{-1}$ (Fixsen et al. 1994). The low frequency band 
(2.2 to 20 cm$^{-1})$ was designed to study the 
CMB and the high frequency band (20 to 96 cm$^{-1}$) 
to measure the 
dust emission spectrum in the Galaxy. We use the 
so-called LLSS (Left Low Short Slow) and RHSS (Right High Short Slow) data 
which cover the low and high frequency bands respectively
(see the FIRAS explanatory supplement).\\
Since in our study we combine the FIRAS and DIRBE data at 140 and $240\mic$,
we have first to check their
consistency. For that, we convolve the FIRAS spectra inside the two DIRBE bands 
at 140 and $240\mic$ and the DIRBE data (at 140 and $240\mic$) with the FIRAS Point Spread Function 
(PSF). 
The PSF is not precisely known for all wavelengths, so we use the approximation 
suggested by Mather (private communication) 
of a $7\degr$ diameter circle convolved with a line of $3\degr$ length
perpendicular to the ecliptic plane (Mather et al. 1986).
 The pixel to pixel correlation shows that the difference of
responsivity between the two instruments is $\sim$$0.1\%$ at $140\mic$ and 
$\sim$$1$$\%$ at $240 \mic$. \\
Before studying the Galactic emission, we have subtracted the CMB and its dipole 
emission from the FIRAS data using the parameters given by 
Mather et al. (1994) and Fixsen et al. (1994).
To remove the interplanetary dust emission, we consider the $25\mic$ map as
a spatial template for the interplanetary dust emission. For the 60 and $100\mic$
maps, we use the zodiacal emission ratios given by Boulanger et al. (1996).
At longer wavelength, we follow
Reach et al. (1995) by taking a zodiacal spectrum $I_{\nu}\propto\nu ^3$.\\ 
To analyse the Galactic emission, it is necessary to choose a reference zero 
emission level because the sky brightness may contain an
isotropic component (Puget et al. 1996).
Following Puget et al., we choose as a reference the extrapolated value 
for zero HI column density of the FIR-HI correlation.
We found an extrapolated emission of
0.65 MJy/sr and 0.56 MJy/sr for the DIRBE 240 and $140\mic$ data (convolved with 
the FIRAS PSF). 
These values are lower than the minimum emission measured in the maps
(equal to $1.70 \pm0.07$ and $1.67 \pm 0.1$ MJy/sr at 240 and $140 \mic$ respectively), 
which shows that all DIRBE pixels (at $7\degr$ resolution) 
contain a significant Galactic contribution.
For the analysis of the large grain temperatures, in Sect. 3, we subtract from the DIRBE data 
at 240 and $140\mic$
0.65 and 0.56 MJy/sr respectively. From the FIRAS data, we subtract the zero HI column density spectrum 
of Puget et al. (1996) smoothed to the resolution of 5.7 cm$^{-1}$.
It is important to check that the reference flux is consistent between DIRBE and FIRAS.
For this, we compare the DIRBE and FIRAS data at high Galactic latitude at $240\mic$
and find an
offset I(DIRBE)-I(FIRAS) equal to -0.15 MJy/sr.
The DIRBE reference zero emission level at $240 \mic$
corrected for this difference  
 is in perfect agreement with the residual spectrum of Puget et al. (1996).
The brightness values subtracted from the DIRBE data should not be used as an estimate of the
CFIBR since it may contain some Galactic emission that is associated with the diffuse ionised
gas. These brightness values are lower than the upper limits on the CFIBR (1.76 and 2.66 MJy/sr
 at 240 and $140\mic$ respectively) derived by Hauser (1995).

\section{\label{map}Large grain temperature maps } 
In this section, to study the spatial variations of the temperature, 
we assume that each individual spectrum can 
be represented as a single modified Planck curve 
with a $\nu ^2$ emissivity law. Obviously the hypothesis of a single temperature
is not valid in molecular regions and near the Galactic plane where 
several clouds with different temperatures overlap. The
temperatures of large grains are
computed from the ratio of DIRBE emissions at 240 and $140 \mic$ (taking into 
account the bandpass) at the FIRAS resolution.
 The very cold component ($T\sim5$ K) if it exists does not
significantly contribute to these bands. 
 We have not used the DIRBE data at shorter wavelengths (60 and $100\mic$) 
 since (1) small grains contribute significantly to the emission 
(D\'esert et al. 1990, Sodroski et al. 1994, Laureijs et al., 1996) and 
(2) the reference zero emission level is difficult to 
evaluate because of the residual zodiacal emission.\\

Fig. 1 shows the temperature map. Uncertainties are about 5 mK
for $|b|<10\degr$, 50 mK for $10\degr<|b|<30\degr$ and 300 mK at higher 
Galactic latitudes. These errors are estimated using
uncertainties in the DIRBE $7\degr$ flux obtained by convolving the rms 
$40'$ DIRBE maps with the FIRAS PSF.

%----------------------------------Fig 1 
\begin{figure*}  
\epsfxsize=16.cm
\epsfysize=18.cm
\hspace{2.cm}
\vspace{-2.cm}
\epsfbox{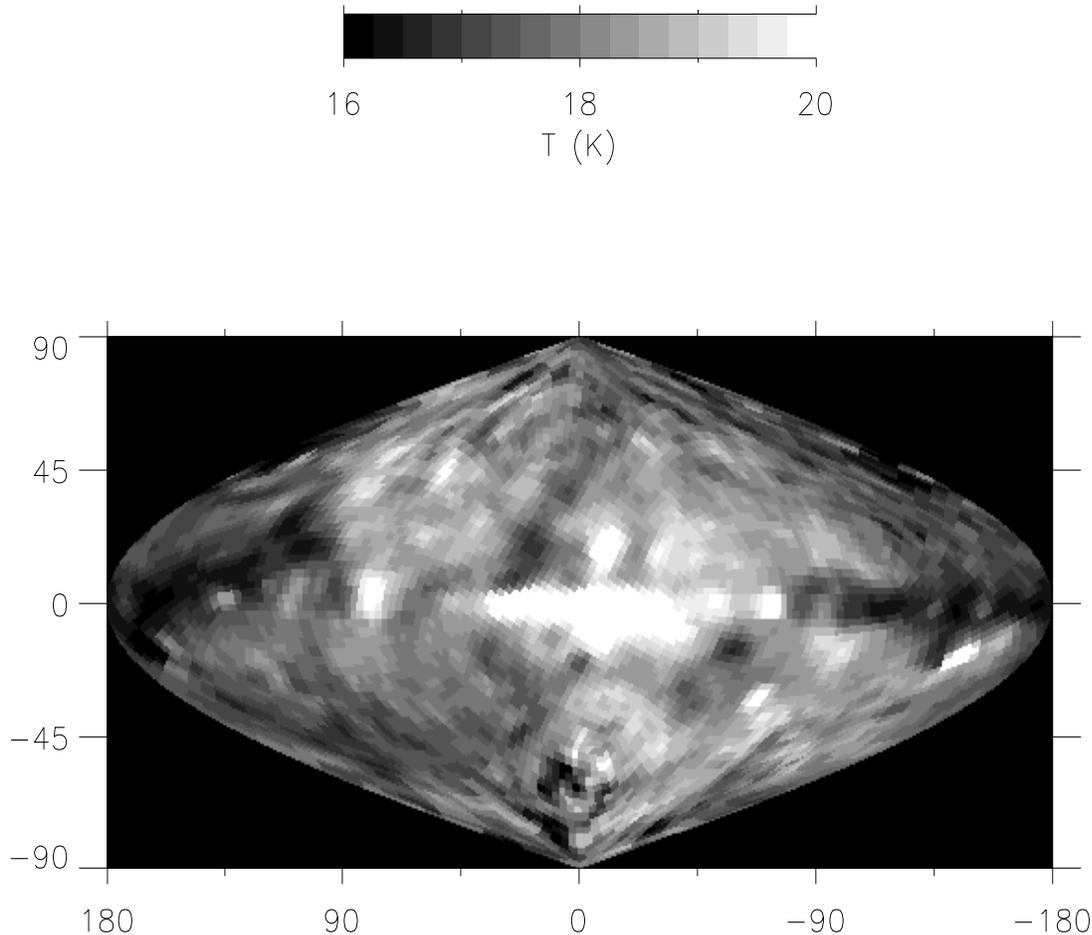}
\caption{Temperature map of the large 
grain dust component computed from the DIRBE maps at 240 and 
$140\mic$ convolved with the FIRAS PSF (in Galactic coordinates).} 
\end{figure*}

Most of the variations visible on Fig. 1 are real. Star forming molecular 
regions (e.g. Orion and Ophiucus) have a higher temperature 
due to dust heated by embedded young stars, while the temperature of clouds 
forming low mass stars is lower (Taurus, part of Cygnus, Chamaeleon, 
Polaris...). Cold spots in the south and north polar caps are not significant
since they correspond to regions with very low emission in the 140 and 
$240\mic$ maps. 
 Cuts along the Galactic plane confirm the decrease of the temperature with increasing 
galactocentric distance already studied by Sodroski et al. (1994).

The scatter diagram presented in Fig. 2 does not show any systematic 
variation of the temperatures with the Galactic 
latitude, which suggests a relatively constant InterStellar Radiation Field (ISRF)
 on large angular scales. 
At smaller scales ($\sim7\degr$), temperature variations at high latitudes (outside large
molecular complexes) indicate
a variation of the ISRF smaller than $\sim30\%$.

It is also possible to build a map of large grain temperature with the
FIRAS data by fitting each spectrum with a modified Planck curve. 
The signal to noise ratio of individual
spectra is low and  temperature uncertainties are typically
two times higher than the DIRBE ones. However, differences of temperatures
obtained from these two independent datasets are within the uncertainties.

%----------------------------------Fig 2 
\begin{figure}  
\epsfxsize=9.cm
\epsfysize=8.cm
\hspace{-0.0cm}
\vspace{-0.5cm}
\epsfbox{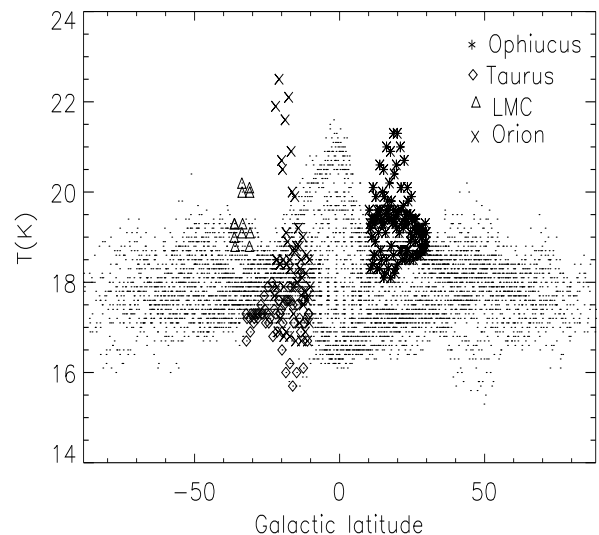}
\caption{ Temperature of the large grains (see Fig. 1) as a function of the
Galactic latitude.
Pixels containing known complexes are identified using different 
symbols: Taurus, Auriga and Perseus ($\diamond$), Orion ($\times$), 
Large Magellanic Cloud ($\triangle$), Ophiucus($\ast$)}
\end{figure}

%----------------------------------Fig 3 
\begin{figure*}  
\epsfxsize=16.cm
\epsfysize=18.cm
\hspace{2.cm}
\vspace{-2.cm}
\epsfbox{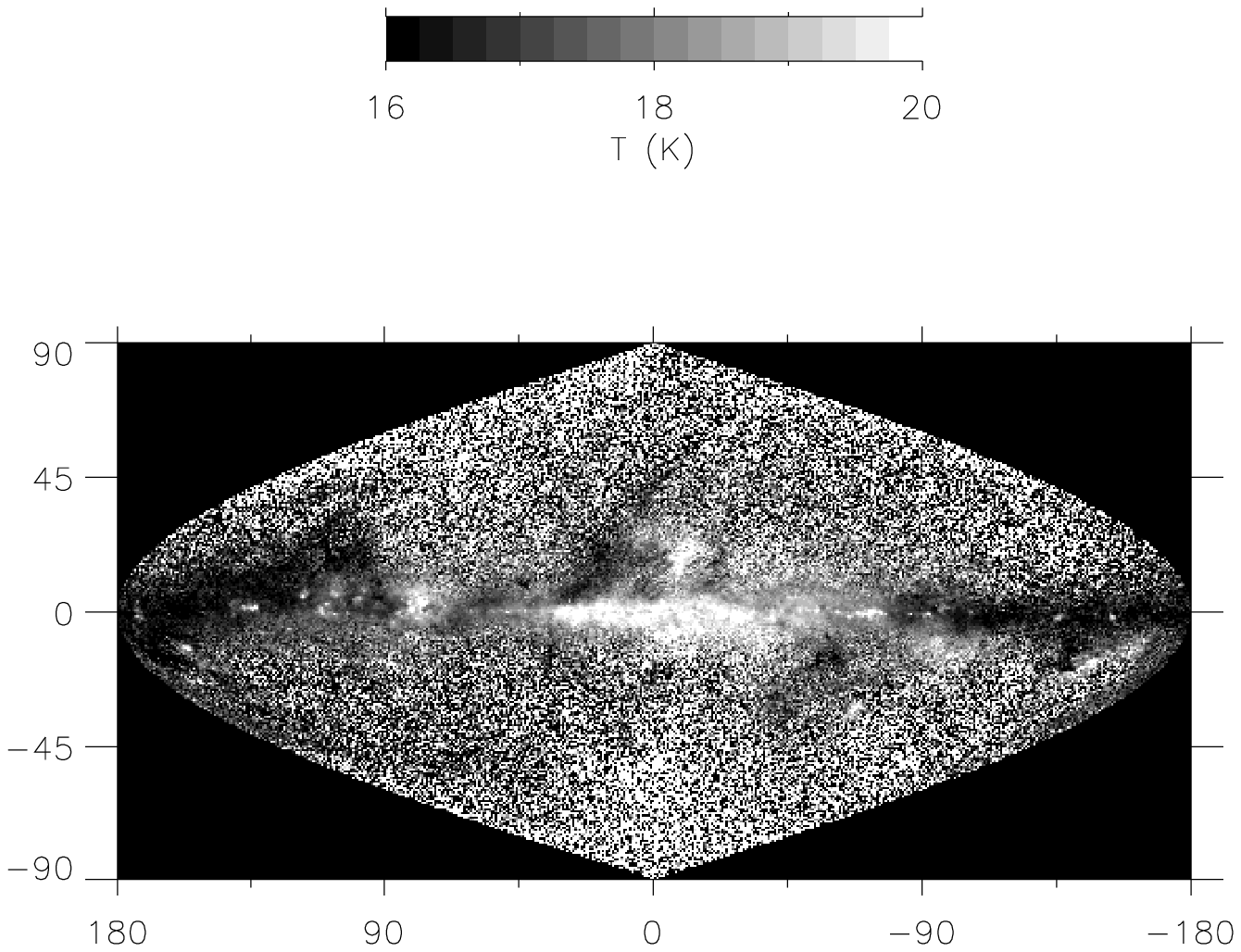}
\caption{ Large grain temperature map at $0.5\degr$ resolution
computed from the DIRBE maps at 240 and $140\mic$ without any convolution
(in Galactic coordinates)}
\end{figure*}

The ISM is known to contain structures over a very wide range of scales with inhomogeneous
physical conditions. Temperatures presented in Fig. 1 correspond to values averaged 
along the lines of sight and within
the $7\degr$ FIRAS beam. For pixels with a
high signal to noise ratio (especially at $|b|<30\degr$), the
DIRBE data without any convolution allow the visualisation of spatial 
variations of the temperatures at a resolution of 40' (Fig. 3). 
For the Taurus, Auriga and Perseus 
(TAP) region, the $^{12}$CO ($J=0-1$) integrated emission
map obtained by Ungerechts \& Thaddeus (1987) at a resolution of $0.5\degr$
is compared in Figs. 4a-b to the temperature map at the DIRBE resolution (projected 
upon the $^{12}$CO grid). 
We see that, 
within this field, there are significant temperature variations (Fig 4b). 
The regions of cold temperatures spatially correlate with the $^{12}$CO filaments.
The highest temperature in Perseus ($\alpha=3h40m10.6s$, $\delta=31d58m7.9s$) 
corresponds to the IC 348 HII region (also known as the Per OB2 cloud).

\section{Cold emission maps}

\subsection{Maps of FIR excess with respect to the 60$\mic$ emission}

Studies based on IRAS images have revealed,
for dense clouds, sharp increases of $I_{\nu}(100)$ without any counterpart of
$I_{\nu}(60)$ (Laureijs et al. 1991 and Abergel et al. 1994).
It is illustrated for the TAP region Figs. 4c-d. 
Image of the component of the dust emission in excess at 100$\mic$
(Fig. 4e) is computed using   
$I_{\nu}(100)_{excess}$=$I_{\nu}(100)$-$I_{\nu}(60)$x$R(100,60)$, where R(100,60) is the 
$I_{\nu}(100)$$/$$I_{\nu}(60)$ ratio assumed constant for the neighbouring emission
("cirrus" component). The emission in excess has been found to be correlated 
with the emission
of dense gas in $^{13}$CO. In this section, we extend the IRAS excess component analysis 
to longer wavelenghts using the DIRBE data.

Let $I_{D}(\lambda)$ ($\lambda=100, 140, 240\mic$) be the DIRBE emission
at 100, 140 and $240\mic$ respectively,
and R($\lambda$,60) the $I_{\nu}(\lambda)$$/$$I_{\nu}(60)$ flux ratio 
observed in "cirrus" clouds.
The excess maps are computed at each wavelength according
to the relationship:
$I_{D}(\lambda)_{excess}=I_{D}(\lambda)-R(\lambda,60)*I_{D}(60)$.

The $I_{D}(\lambda)$ vs $I_{D}(60)$ correlation diagrams 
(used to derive $R(\lambda,60)$)
present a high dispersion due
to real color variations across the sky (due
to temperature effects and/or abundance variations). 
However, the signal to noise ratio and the angular resolution of DIRBE data
do not allow a proper determination of 
the spatial variations of $R(\lambda,60)$
 over the whole sky. Therefore, we have decided to work with averaged
values estimated by eye-fitting the $I_{D}(\lambda)$ 
vs $I_{D}(60)$ correlation diagrams for all pixels at $|b|>20\degr$ and 
$|\beta|>20\degr$, 
outside known molecular complexes and the Magellanic Clouds.
 These values are equal to 
$4\pm0.7$, $6\pm1$ and $3.2\pm0.5$ at 240, 140 and $100\mic$ respectively. 
The error bars are due to the dispersion of the points. 

%----------------------------------Fig 4 
\begin{figure*}
\epsfxsize=22.cm
\epsfysize=28.cm
\hspace{-4.cm}
\vspace{-4.6cm}
\epsfbox{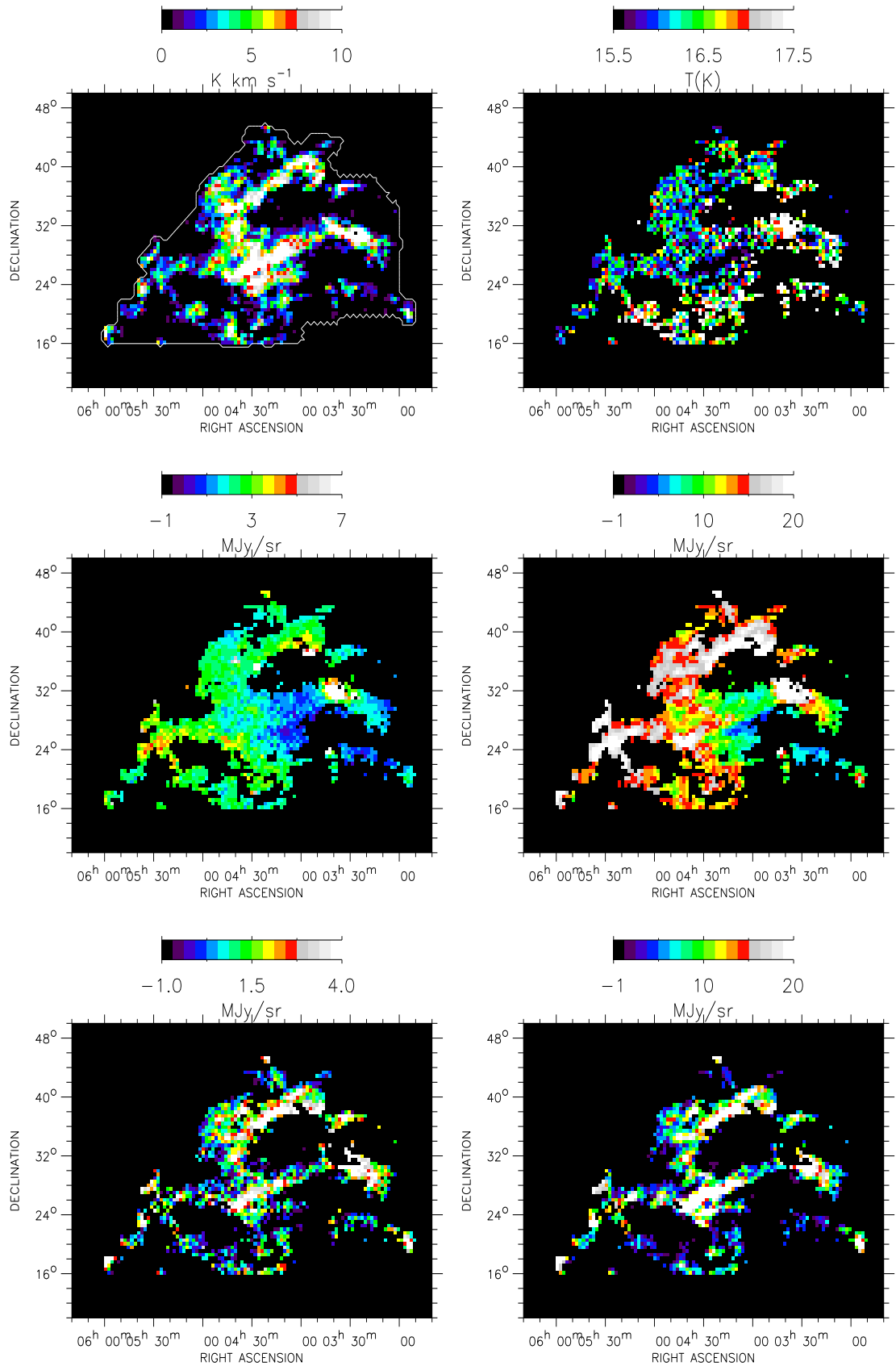}
\caption{Taurus, Perseus and Auriga region:
(a) CO J=0-1 integrated emission from Ungerechts \& Thaddeus (1987) 
(b) large grain temperature map (the same as Fig. 3
projected on the CO grid) (c) $60\mic$ and (d) $100\mic$ emission.
Cold component of the dust emission at 100 (e) and $240\mic$ (f) 
corresponds to bright features. They are due to a sharp increase
of $I_{D}(100)$ and $I_{D}(240)$ without any counterpart of $I_{D}(60)$.
These features are very well correlated with the CO emission, 
confirming that, in this region, molecular emitting regions coincide with cold regions
(Abergel et al. 1994)}
\end{figure*}

Using these three color ratios, we
can also compute R(140,100) and R(240,100) (see Table 1).
They are remarkably close
to the slope of the 
$I_{D}(\lambda)$ vs $I_{D}(100)$ correlation diagrams in
Dwek et al. (1997). The discrepancy observed for R(100,60) is due to the 
fact that we do not study the same part of the sky ($|b|>30\degr$ and 
$|\beta|>40\degr$ for Dwek et al.). 
With the same data points as ours and the same data preparation as Dwek, 
a value 
of 3.4 is obtained (Arendt, private communication), which is very close to
our value of 3.2.
Our color ratios differ significantly from those of Boulanger et al. (1996), 
especially R(240,100).
We have checked that these differences cannot be accounted for by the fact that we do not study
the same part of the sky. The color ratios depend on which map is used as independent variable 
for the correlation (HI for Boulanger et al., $100\mic$ for Dwek et al. and $60\mic$ 
in the present study). 

%----------------------------------Table 1 
\begin{table*} \caption{\label{tab1}FIR and submm color ratios for the
whole sky at high Galactic latitude}
\begin{flushleft} 
 \begin{tabular}{|l|c|c|c|c|c|} \hline        
Publications & R(100,60) & R(140,60) & R(240,60) & 
R(140,100) & R(240,100) \\ \hline
This paper (1) & $3.2\pm0.5$ & $6\pm1$ & $4\pm0.7$ & 
$1.9\pm0.6$ & $1.25\pm0.4$   \\ \hline
Dwek et al. (1997) (2) & 6.25 & & & 1.93 & 1.28  \\ \hline
Boulanger et al. (1996) (3) & & & & 2.36 & 2.02 \\ \hline
\end{tabular}\\
(1) Ratios are estimated via the $60\mic$ emission (for $|b|>20\degr$ and
$|\beta|>20\degr$)\\
(2) Ratios are estimated via the $100\mic$ emission (for $|b|>30\degr$ 
and $|\beta|>40\degr$)\\
(3) Ratios are estimated via the HI emission (for $|b|>20\degr$)\\
\end{flushleft} \end{table*} 

We have used our estimates of $R(\lambda,60)$ (given in Table 1) 
to compute the excess component maps at 
100, 140 and $240\mic$. These maps still contain a zodiacal residual emission 
(mainly coming from the $60\mic$ map) visible on large scale. 
It is difficult to produce all sky maps so we decide to concentrate our
study on local features. Low frequency structures in excess maps are removed using 
a $21\degr$x$21\degr$ median filter.
The filtering also corrects for the large scale variations of the color ratios
$R(\lambda,60)$ and for the isotropic contributions (as the CFIBR).

\subsection{Temperature of the excess component in the 
Taurus, Auriga and Perseus region}

The excess maps at 100 and $240\mic$ in the TAP region are presented 
Figs. 4 e-f. We have chosen 8 lines of sight to determine the 
temperature of the dust associated with the filaments in excess 
(Table 2, T$_{1}$).
It is clear that these filaments correspond to cold regions with
temperatures around 13 K.

To test the effect of spatial variations from cloud to cloud 
in the color ratios R($\lambda$,60), we
have derived the temperatures using excess maps computed with local
color ratios derived in the vincinity of the TAP region:
R(240,60)=8, R(140,60)=10 and R(100,60)=5. 
The new temperatures we obtain are
very close to the ones determined with the averaged color ratios
(Table 2, T$_{1}$ and T$_{2}$). We see that variations 
in the color ratios do not affect the determination of the 
temperatures by more than 0.2 K.

Finally, to check whether the temperature determination may be biased by the 60$\mic$
removal, we compute the temperature using an "ON-OFF" method:
a background emission is estimated for each pixel of the filaments at a distance of
typically $3\degr$ (OFF position). This
 background is locally removed from the emission.
Temperatures derived with this method
are very close to the ones derived with the excess maps (Table 2).
Therefore, we conclude that all
bright filaments in the excess maps of the TAP region definitively
correspond to cold filaments. We will see in Sect. 5 that we can generalise this result
for the whole sky at $|b| > 10\degr$. So, in the following, the excess maps are called
``cold emission maps''.

 %----------------------------------Table 2 
\begin{table*} \caption{\label{tab1}Temperatures of the 
bright filaments in the TAP region determined using (1) the excess
maps derived with the averaged color ratios (Table 1 and Sect. 4.1), (2)
the excess maps derived with local color ratios computed in the
TAP region and (3) an "ON-OFF" method.}
\begin{flushleft} 
 \begin{tabular}{|l|l|c|c|c|} \hline        
RA (1950) & DEC (1950) & T$_{1}$ (1) & T$_{2}$  (2) & 
 T$_{3}$ (3) \\ \hline
%
%(451,190)
04h03m33.64s   & 40d24m36.8s  & 14.6$\pm$0.3 & 14.6$\pm$0.3 & 
15.3$\pm$0.6 \\  
%
%(452,152) 
04 13 11.65 & 28 43 12.6  & 12.9$\pm$0.2 & 13.0$\pm$0.2 & 
13.2$\pm$0.2 \\ 
%
%(445,181)
04 16 15.12   & 38 10 35.1  & 14.0$\pm$0.2 & 13.9$\pm$0.2 & 
14.3$\pm$0.2 \\ 
%
%(443,178)
04 20 20.24   & 37 24 39.3  & 13.5$\pm$0.2 & 13.3$\pm$0.3 & 
14.9$\pm$0.2 \\  
%
%(446,146)
04 23 41.15 & 27 20 1.8  & 13.9$\pm$0.3 & 13.9$\pm$0.3 & 
 14.9$\pm$0.3 \\ 
%
%(445,137)
04 27 6.25   & 24 40 18.0  & 12.6$\pm$0.3 & 12.5$\pm$0.3 & 
13.7$\pm$0.4\\ 
%
%(445,136) 
04 27 18.82  & 24 22 7.6  & 12.5$\pm$0.2 & 12.4$\pm$0.2 & 
13.3$\pm$0.3\\ 
%
%(439,141) 
04 35 9.36   & 26 14 4.6  & 13.2$\pm$0.3 & 13.0$\pm$0.2 & 
13.5$\pm$0.2 \\ \hline 
\end{tabular}\\
\end{flushleft} \end{table*} 

\subsection{Correlation of the cold component with molecular gas}

%----------------------------------Fig 5 
\begin{figure*} 
\epsfbox{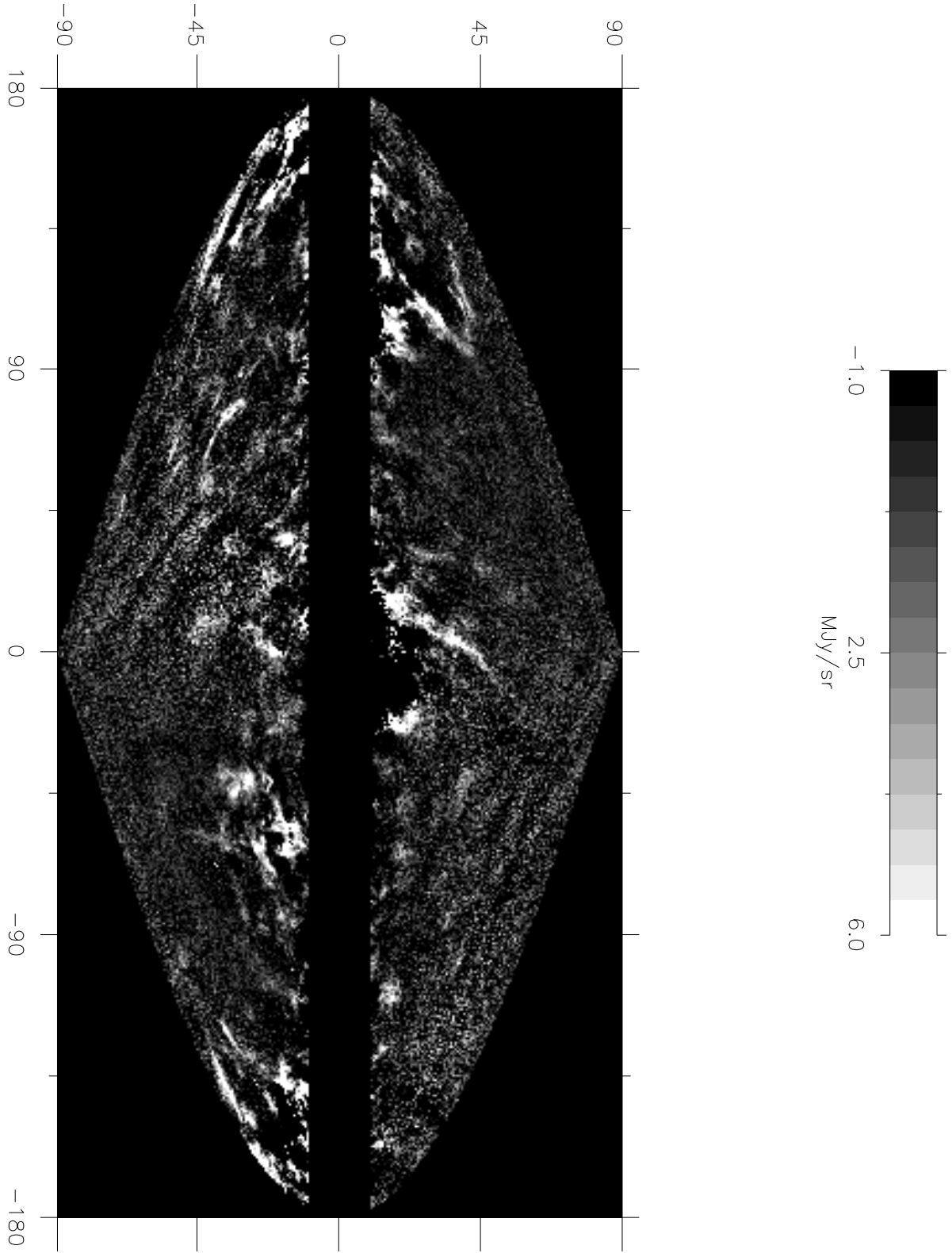}
\hspace{-2.cm}
\caption{Cold component of the dust emission at $240 \mic$, 
$I_{D}(240)_{cold}=I_{D}(240)-R(240,60)*I_{D}(60)$,
where $R(240,60)$ is obtained by computing the slope of the $I_{D}(240)$ 
vs $I_{D}(60)$ correlation diagrams for all pixels at $|b|>20\degr$ and 
$|\beta|>20\degr$, 
outside large molecular complexes and Magellanic Clouds (see the text). 
We have removed the data at $|b|\le10\degr$ because of high spatial variations of
$R(240,60)$.}
\end{figure*}

The whole sky map of the cold component at $240\mic$ is presented in Fig. 5.
We exclude in our analysis the regions at $|b|\le10\degr$ because 
of high spatial variations of $R(\lambda,60)$.
All bright features at $|b|>10\degr$ are associated with large molecular complexes:
Polaris, Camelopardalis and Ursa Major ($b\sim30\degr$, $l\sim140\degr$), 
Chamaeleon ($b\sim-16\degr$, $l\sim300\degr$),
part of Orion ($b\sim-15\degr$, $l\sim210\degr$) and Taurus ($b\sim-20\degr$, $l\sim180\degr$) 
as illustrated in more details in Figs. 4e-f. The network of 
 bright filaments at negative latitudes around $b\sim-35\degr$ and $l\sim90\degr$
is associated with molecular emission
near the OB association I Lacerta. However, not all molecular complexes are 
visible on Fig. 5. For example, most of the $\rho$ Ophiuchus complex
is not detected, since it contains several B stars and a large number of embedded young stellar objects. 
 We can also compare the 240$\mic$ cold map to the map of the residuals of 
the IR/HI
correlation (Boulanger et al. 1995). All structures in the cold maps appear 
as positive excess in the IR/HI residuals, but the reverse is not true.
Locally heated molecular clouds and
ionised clouds are not present in Fig. 5 (as the one around 
the nearby $B1$ star Spica located at $l=316\degr$, $b=51\degr$, Zagury et al., 1998).

%----------------------------------Fig 6 
\begin{figure}  
\epsfxsize=9.cm
\epsfysize=8.cm
%\hspace{2.cm}
\vspace{0.0cm}
\epsfbox{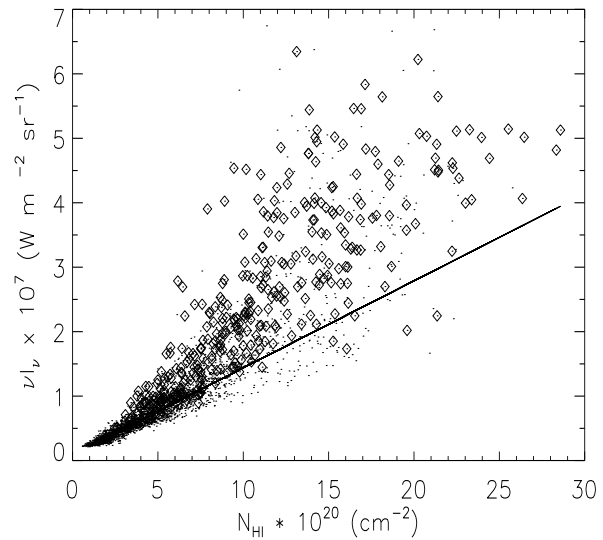}
\caption{Correlation diagram between the DIRBE $240\mic$ and the HI emissions convolved
in the FIRAS beam. The data 
points for which the cold component is detectable at $240\mic$ 
are surrounded.
The continuous line, from Boulanger et al. (1996), represents the fit 
to data at N$_{HI}$$\le$4.5 $10^{20}$ H cm$^{-2}$}
\end{figure}

Boulanger et al. (1996) have correlated the FIR emission from dust as measured by
COBE at high $|b|$ with the 21-cm emission from gas as measured by the Leiden/Dwingeloo 
survey of HI (Hartmann \& Burton, 1997). At all wavelenghts ($\lambda\ge100\mic$), 
the correlation is tight for N$_{HI}$$\le$5 $10^{20}$ H cm$^{-2}$.
For higher N$_{HI}$, the data points depart
from the low emission correlation. 
They interpret the change in the slope in the FIR-HI correlation 
as an increasing contribution of molecular gas for N$_{HI}$ 
larger than 5 $10^{20}$ H cm$^{-2}$.
The $240\mic$-HI correlation at the FIRAS resolution is shown in Fig. 6 for $|b|>10\degr$, using 
the same HI dataset. 
We have surrounded all data points for which the cold component is detectable 
(see Sect. 5.1)
at $240\mic$.  
Most of these "cold pixels" depart from the low correlation emission 
(continuous line of Fig. 6 from Boulanger et al., 1996) because of the increasing
contribution of cold molecular regions. The other points which also 
depart from the low HI column density correlation correspond
to emissions coming from (1) HII regions, (2) warm molecular clouds and (3)
cold molecular clouds too small to be detected as such
in the FIRAS beam.
The minimum HI column density for the "cold pixels" at $240\mic$
is 2.5 $10^{20}$ H cm$^{-2}$. We consider this value as a threshold for the detection
of cold emission in DIRBE data convolved in the FIRAS beam.

\section{FIRAS spectra of the warm and cold components}
In this section, we combine the DIRBE and FIRAS data to 
extend the study of the Galactic emission (especially the separation between the
warm and cold components) to longer wavelenghts. Thus, we work at the FIRAS
resolution of $7\degr$.

\subsection{Temperature and optical depth of the cold and warm components}
We convolve the cold DIRBE emission maps 
with the FIRAS PSF ($I_{F}(240)_{cold}$, $I_{F}(140)_{cold}$ and 
$I_{F}(100)_{cold}$). These maps contain the emission of the cold dust 
component in the FIRAS beam. 
Even with the low FIRAS resolution, pixels for which the cold component
is significant can be identified.
To select these pixels, we compute the root mean square ($\sigma_{\lambda}$)
for each cold map
by fitting the brightness histograms by a gaussian curve. 
We find $\sigma_{240}$=0.41 MJy/sr,
$\sigma_{140}$=0.54 MJy/sr and $\sigma_{100}$=0.22 MJy/sr.
The dominant sources of noise are spatial variations in $R(\lambda,60)$.
 We restrict our analysis to $|b|>10\degr$ and separate the data in two sets
 with and without significant cold emission. The pixels with significant
 cold emission are defined with:
$I_{F}(\lambda)_{cold}$$>$$3\sigma_{\lambda}$ at 100, 140 and $240\mic$.
Mean cold emissions are equal to 1.0, 3.5 and 3.9 MJy/sr at
100, 140 and $240\mic$ respectively.\\
The spectra of pixels with no significant cold emission ($61\%$ of the sky) are
fitted by a single temperature Planck curve
with a $\nu^2$ emissivity law. The mean temperature for this set of pixels
is equal to 17.5 K with a dispersion of 2.5 K. The high dispersion comes from low
signal to noise ratio of individual FIRAS spectra at high Galactic latitude.
Therefore, we confirm that 
the averaged temperature of the interstellar dust in the atomic medium
is rather uniform and of about 17.5 K (D\'esert et al. 1990, 
Boulanger et al. 1996).  \\

For pixels with significant cold component (206 pixels which represent
3.4$\%$ of the sky),
 we measure the optical depth
and temperature of the warm and cold components.
Let $I_{F}(\lambda)$ be the intensity of the DIRBE emissions at the FIRAS resolution,
$I_{F}(\lambda)_{cirrus}$ the intensity
of the "cirrus" contribution (which represents the HI clouds and warm molecular envelops) 
and $I_{F}(\lambda)_{cold}$ the intensity
of the cold dust contribution.
We compute $I_{F}(\lambda)_{cirrus}$ using:\\
$I_{F}(\lambda)_{cirrus}$ = $I_{F}(\lambda)$ - $I_{F}(\lambda)_{cold}$\\
The temperature and optical depth of the 
"cirrus" component ($T_{cirrus}$ and $\tau_{cirrus}$) are computed using 
$I_{F}(240)_{cirrus}$/$I_{F}(140)_{cirrus}$ and 
assuming a $\nu^2$ emissivity law. Then, cold component spectra are obtained by 
removing the "cirrus" modified Planck curves, defined by $T_{cirrus}$ and $\tau_{cirrus}$,  
from the FIRAS spectra. Each of these cold component spectra is fitted 
 by a modified Planck curve with a $\nu^2$ emissivity law 
(the $C^{+}$ line at $158\mic$ is removed 
before doing the fits). We have checked that a free $\alpha$ does not significantly
decrease the $\chi$$^{2}$ value of the cold component fits. As an illustration,
fits of the cold component spectrum at $l=-305.2\degr$ and $b=-15.75\degr$
with (a) T=19 K, 
$\tau=1.6 10^{-7}$, $\alpha$=1.4 and (b) T=15.25 K, 
$\tau=3.9 10^{-8}$, $\alpha$=2 give the same $\chi$$^{2}$ value of 1.26.
We see that the poor signal to noise ratio does not 
allow to constrain $\alpha$.
Moreover, the signal to noise 
ratio of individual cold spectra is too low for about $64\%$ of them to even determine
properly the temperature and optical depth assuming $\alpha$=2. 

%----------------------------------Fig 7 
\begin{figure}  
\epsfxsize=9.cm
\epsfysize=8.cm
%\hspace{2.cm}
\vspace{0.0cm}
\epsfbox{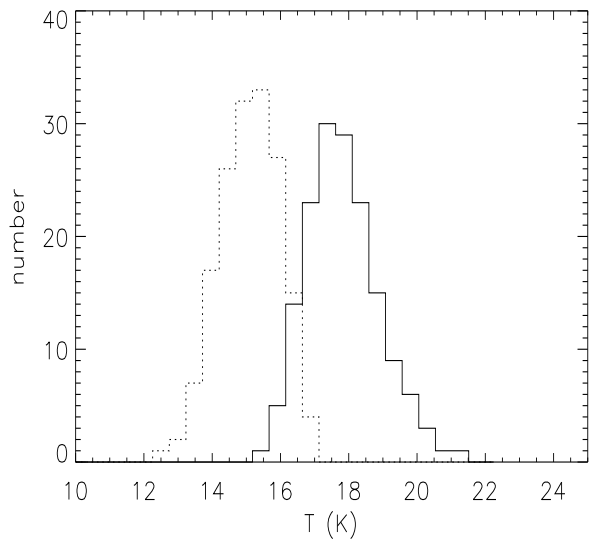}
\caption{Histograms of the temperatures for the "cirrus" component 
(continuous line) and the cold component (dashed line), for pixels containing 
significant cold emission (see the text)}
\end{figure}

The signal to noise ratio is larger with DIRBE than with FIRAS data. Thus,
we use the DIRBE 100, 140 and $240\mic$ cold maps to compute
the temperature $T_{cold}$ and optical depth $\tau_{cold}$ of 
the cold component. We assume a $\nu^2$ emissivity law since 
the emissivity
index ($\alpha$) cannot be properly constrained. This $\nu^2$ emissivity law corresponds to standard 
interstellar dust grains (Draine \& Lee, 1984).
The cold component has optical depths (normalised to 1 
cm$^{-1}$) distributed around 5.3 $10^{-8}$ with a dispersion of 4.1 $10^{-8}$
and temperatures distributed around 15 K with a dispersion 
of 0.8 K (Fig. 7). The coldest temperatures ($\sim$13 K) are located in the Taurus region.  
Then, we derive spectra of the dust outside 
cold regions ("cirrus" component) by subtracting the cold modified Planck 
curve defined by $T_{cold}$ and $\tau_{cold}$ from the FIRAS spectra. Finally, 
each "cirrus" spectrum is individually fitted in order to determine 
$T_{cirrus}$ and $\tau_{cirrus}$. 
The "cirrus" component has optical depths distributed around 9.8 $10^{-8}$ with 
a dispersion of 6.1 $10^{-8}$ and temperatures distributed around 17.8 K 
with a dispersion of 1.2 K (Fig. 7).
These temperatures are significantly different from the cold 
temperatures determined for the same set of pixels.
An example of the decomposition in two components 
is shown in Fig. 8 for one pixel located in
the Chamaeleon complex ($l=305.2\degr$ and $b=-15.75\degr$). 

%----------------------------------Fig 8 
\begin{figure}  
\epsfxsize=9.cm
\epsfysize=8.cm
%\hspace{2.cm}
\vspace{0.0cm}
\epsfbox{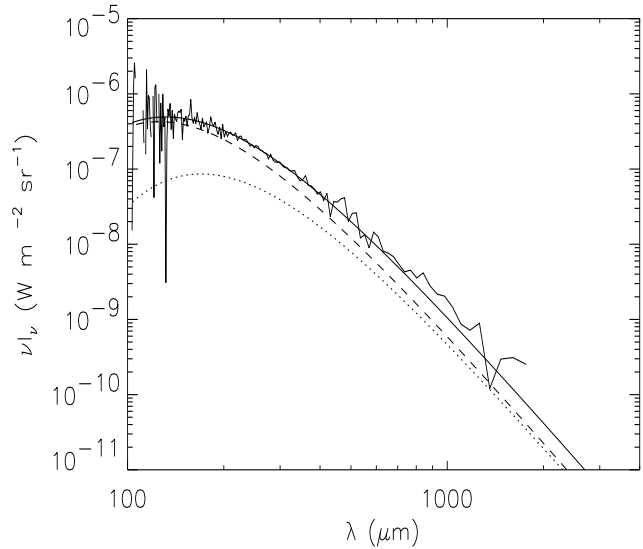}
\caption{Example of the decomposition into the "cold" and "cirrus" components 
for one selected pixel ($l=-305.2\degr$ and $b=-15.75\degr$) located in the 
Chamaeleon cloud.
The parameters of the modified Planck curves are:
$T_{cold}$=14.2 K, $\tau_{cold}$=6.9 $10^{-7}$ (dotted line) and $T_{cirrus}$=19.2 K,   
$\tau_{cirrus}$=5.5 $10^{-8}$ (dashed line). 
The solid line corresponds to the synthetic total emission.}
\end{figure}

\subsection{Residual emission}
The presence of very cold dust (T=4-7 K), not detectable using the DIRBE data at 
140 and 240 $\mic$, may be revealed on the residual FIRAS emission.
We compute two residual spectra: one for regions with detected cold emission, 
by removing the cirrus and cold contributions from the FIRAS spectra, 
and one for regions with no detected cold emission, by removing only the cirrus 
contribution from the FIRAS spectra.
In Fig. 9, we present the mean emission spectra of the warm and cold
components together with the mean residual spectrum
for regions with detected cold emission.
The mean residue  is negative in the range 
$280-390\mic$ (2$\%$ of the total mean emission) and positive above $400\mic$
($\sim$15$\%$ of the total mean emission at $600\mic$).
We have checked that this pattern persists if we fit the spectra with
a fixed emissivity index different than 2.

%----------------------------------Fig 9 
\begin{figure}  
\epsfxsize=9.cm
\epsfysize=8.cm
%\hspace{2.cm}
\vspace{0.0cm}
\epsfbox{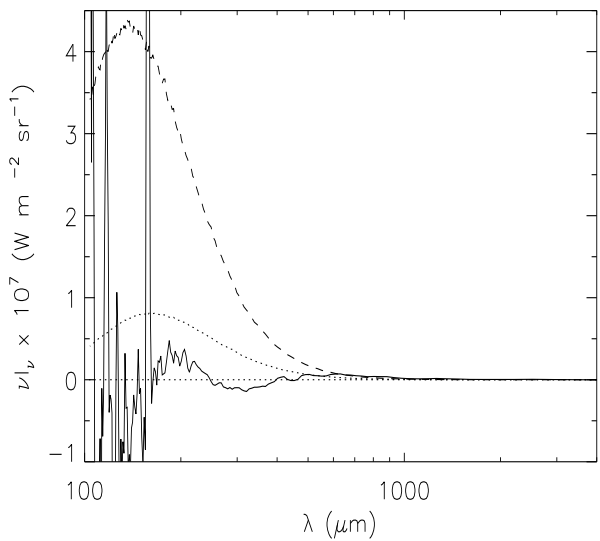}
\caption{Mean emission spectra of the warm (dashed line) and cold (dotted line)
component for regions with detected cold emission. 
The continuous line shows the mean residual spectrum (multiplied by 5), also for
regions with detected cold emission}
\end{figure}

The residual to the one temperature fit of spectra without cold emission
shows a similar pattern, comparable to the $1\sigma$ positive submm excess of Dwek et
al. (1997). This excess is typically 25 times smaller than the one detected in
Reach et al. (1995). It could be due to very cold dust but
we have checked that a temperature distribution of the Galactic components
qualitatively reproduces
the spectral shape of the residues. 
We have seen (in Sect. 5.1 and Fig. 7) that the temperature of the "cirrus" is $\sim17.5$ K
while the temperature of the cold regions can be as low as 13 K. Obviously, inside the FIRAS 
pixels with detected cold emission, 
there is a range of temperature at least from $T_{cirrus}$ to
$T_{cold}$. Moreover, the cold regions appear to coincide with molecular clouds which
have a typical
size smaller than the FIRAS and DIRBE beams. Therefore, pixels with no detected cold 
emission can also contain small size molecular clouds inducing a range of temperatures.
Our residual emission does not contain any detectable very cold dust.

\section{Discussion}
\subsection{Warm and cold components}

We have used the spectral information of DIRBE and FIRAS to separate the
different emissions of the ISM. The predominant emission of the whole sky in the FIR 
comes from "cirrus" clouds (as defined earlier: HI clouds and 
warm molecular envelopes).
The large grain temperature in "cirrus" is relatively uniform 
and around 17.5 K (Sect. 5.1). 
We do not see any systematic variation of the temperatures with the Galactic 
latitude which suggests
a relatively constant ISRF on very 
large angular scales. 
For a $\nu^2$ emissivity law, the dust temperature scales as the radiation field
intensity to the power 1/6 and the spatial variations
 of temperature can be converted to variation in 
the radiation field intensity smaller than $\sim30\%$.
The cold component of the dust emission with a temperature $\sim15$ K 
is spatially correlated with large molecular
complexes with low star forming activity. The coldest temperature is $\sim13$ K. 
Since these temperatures are averaged
inside the FIRAS beam, the physical value of the dust temperature in 
molecular regions can obviously be lower than 13 K. Such cold temperatures
have also been found by the balloon-borne experiment PRONAOS in
several dense cores in star forming regions with an angular resolution of 2-3.5' (Ristorcelli
et al., 1996, 1998, Serra et al., 1997).\\
 
An important question is to know whether the drop in the 
temperature is only due
to the attenuation of the radiation field in the molecular clouds or whether it
also results from a change in the large grain properties. 
If the small particles disappear by sticking onto large grains within dense gas,
it is natural to speculate that the emissivities of the large grains are
different. The large grains could become fluffy and this will
affect the long wavelength emissivity (e.g Bazell and Dwek, 1990).
The IRAS emission at 12 and $25\mic$ is observed to drop where the
cold emission is present, and this has been interpreted 
as due to a low abundance of PAHs and VSGs (Bernard et al., 1992). 
We find here that low 
abundances are correlated with the drop in the temperatures of large grains.
We derive an averaged value of R(140,240) in cirrus clouds and in cold regions using 
the temperature histograms (Fig. 7). 
In Table 3 we compare these color ratios with those determined
in the Bernard et al. (1992) model which computes with a radiative
transfer code the FIR emission of a non-homogeneous, spherically symmetric
cloud with a standard dust composition. The large error bars of our estimates of
R(140,240), due to the dispersion of the temperature
from clouds to clouds, do not allow to conclude
if the drop in the temperature is only due
to the attenuation of the radiation field or if it also
results from a change in the large grains properties. 
It is necessary to study in detail
individual molecular clouds which is beyond the scope of this paper which presents a
statistical analysis on the whole sky.

%----------------------------------Table 3 
\begin{table} \caption{\label{tab2} Values of the R(140,240) ratio derived
in cirrus clouds and in cold regions with DIRBE data and computed with 
the model developped by Bernard et al. (1992) for the case 
$\beta$=1, central density n$_{H}^{0}$=10$^{4}$ cm$^{-3}$ and A$_{v}$=1, 4, 6, 20. The value obtained 
by D\'esert et al. (1990) in cirrus clouds (A$_{v}$=0) is also reported.} 
\begin{center} 
 \begin{tabular}{|c|c|} \hline        
 & R(140,240) \\ \hline
Cirrus clouds & $1.3\pm0.4$ \\ \hline
Cold regions & $0.9\pm0.4$\\ \hline
A$_{v}$=0 & 1.26 \\ \hline
A$_{v}$=1 & 1.1 \\ \hline
A$_{v}$=4 & 0.77 \\ \hline
A$_{v}$=6 & 0.64 \\ \hline
A$_{v}$=20 & 0.46 \\ \hline
\end{tabular}\\
\end{center} \end{table} 

%----------------------------------Fig 10
\begin{figure}  
\epsfxsize=9.cm
\epsfysize=8.cm
%\hspace{2.cm}
\vspace{-2.cm}
\epsfbox{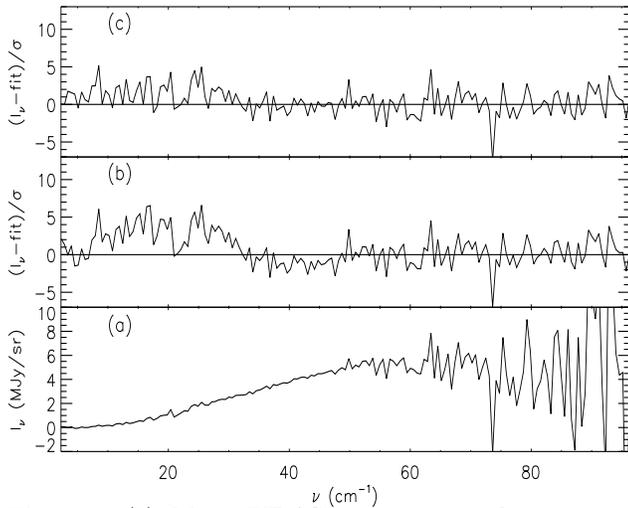}
\caption{(a) Mean FIRAS spectrum in the region with 
30$\degr$$\le$b$\le$60$\degr$ and 180$\degr$$\le$l$\le$270$\degr$.
 (b) Residuals after removing 
a single component modified
black body with $\alpha$=2. (c) Residuals after removing a single
component modified black body with $\alpha$=2 from
the mean spectrum (a) corrected for the reference zero emission spectrum 
(see Sect. 2 for more details). We see that, at high latitude, more than $80\%$
of the submm excess (b) is due to the reference zero emission spectrum.}
\end{figure}

%----------------------------------Fig 11
\begin{figure}  
\epsfxsize=9.cm
\epsfysize=8.cm
%\hspace{2.cm}
\vspace{0.cm}
\epsfbox{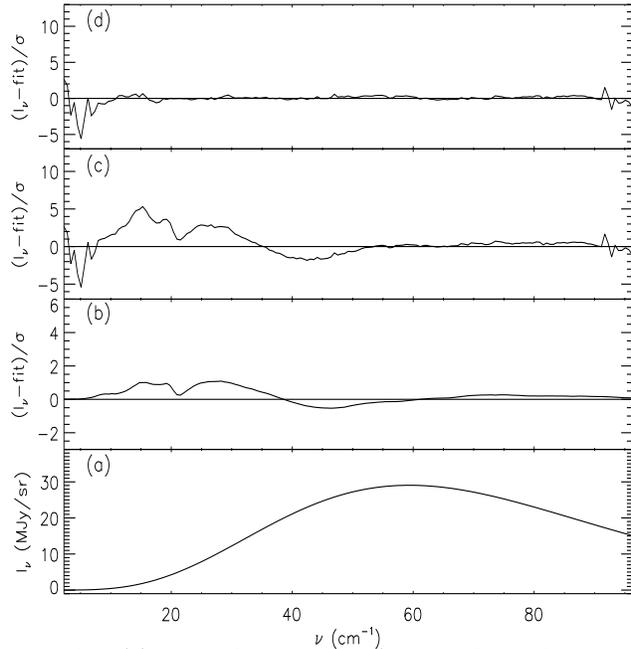}
\caption{(a)Synthetic spectrum (sum of 2 modified Planck curves
with $\alpha$=2, T$_{cold}$=15 K, $\tau_{cold}$=5.3 10$^{-8}$ and 
T$_{cirrus}$=17.8 K, $\tau_{cirrus}$=9.8 10$^{-8}$). (b) 
Residuals after removing a single component modified
black body with $\alpha$=2 from the synthetic spectrum. (c) 
Residuals after removing a single component modified
black body with $\alpha$=2 from the synthetic spectrum added of
the reference zero emission spectrum.
The residual emission 
contains the same submm excess as the one shown in the Fig. 3 of 
Reach et al. (1995) 
(d) Residuals after removing two
components modified black body with $\alpha$=2 from the
synthetic spectrum with
the reference zero emission spectrum added 
(T$_{1}$=17.1 K, $\tau_{1}$=1.5 10$^{-7}$,
T$_{2}$=8.4 K, $\tau_{2}$=8.15 10$^{-8}$). The detection of the
very cold component at 8.4 K is an artefact due to the combination of the cold one and
the reference zero emission spectrum.
The fluctuations in (b,c,d) are due to the standard deviations taken from the data 
(the same as in Reach et al.).}
\end{figure}

\subsection{Very cold dust in the Galaxy ?}
The problem of the presence of 
a very cold component in the Galaxy (T=4-7 K) is very important 
since a positive detection could possibly imply the existence of an unknown 
population of Galactic
grains. Moreover, a very cold dust component at high Galactic latitude would produce
a submm Galactic foreground extremely confusing for the future experiments  
dedicated to study the cosmological backgrounds.
In this section, we compare our results with the previous studies of
 the submm emission of dust from balloon-borne
experiments (Fischer et al., 1995 and Masi et al., 1995) and FIRAS
data (Wright et al., 1991 and Reach et al., 1995).

Fischer et al. and Masi et al. have detected a submm excess (with respect to a $\nu^2$ modified
Planck curve fit) around the star $\mu$ Pegasi and in the Aries and Taurus region
respectively. Their results are strongly constrained by the absolute values of the IRAS $100\mic$ emission.
However, these values are uncertain, since the $100\mic$ calibration of IRAS is known to be different
from that of COBE, and depend on the angular scale of sources.
COBE/IRAS brightness ratio of 0.7 
applies to very large scale emission (Sodroski et al., 1994) but this factor is
close to 1 for angular scales smaller than $1\degr$.
Moreover, we detect a significant cold component with a temperature $\sim15.5$ K 
in the FIRAS maps
in both regions but no very cold dust emission.
We can conclude that the regions mapped by these two balloon borne experiments 
are not representative of the diffuse interstellar medium and
contain submm excess due to cold molecular clouds.

Reach et al. (1995) have fitted FIRAS spectra, averaged in several bins 
over the whole sky and in the Galactic plane, with two components 
and found that the
optical depths of the very cold  component (4-7 K) and 
warm component (16-21 K) are correlated.
This result strongly supports the idea that the submm excess due to the
very cold component has a  Galactic origin.
However, the Reach et al. analysis is based on total power FIRAS 
spectra which could include
a non-Galactic emission component.  
Puget et al. (1996) have found, in the FIRAS data after the removal of
the HI correlated emission, a positive residual emission. They argued 
that most of this emission comes from an isotropic component. At high
latitudes, the presence of such an isotropic component 
can explain the detection of very cold emission with a
two component fit (Fig. 10, to be compared to Fig. 3b of Reach et al.). 

At intermediate latitudes ($10\degr<|b|<30\degr$), the isotropic 
emission becomes negligible
and therefore cannot explain alone the very cold component found 
by Reach et al. However,
we have seen that most of the cold emission ($\sim$15 K) is detected using
DIRBE data
in this part of the sky (Sect. 4 and Fig. 5).
To check whether 
the submm excess of Reach et al. is due to this cold emission, 
we have built a synthetic spectrum (Fig. 11a)
made of two modified Planck curves with temperatures and 
optical depths equal to
the averaged values found for pixels with detected cold emission
(T$_{cold}$=15 K, $\tau$$_{cold}$=5.3 $10^{-8}$, T$_{cirrus}$=17.8 K 
and $\tau$$_{cirrus}$=9.8 $10^{-8}$, see Sect. 5.1). The
residual emission after removing a single component modified black
body shows a
submm excess comparable to that found in Reach et al. (Fig. 11 b).
The addition of the
isotropic component to the synthetic spectrum increases the amplitude of 
this excess (Fig. 11 c). The fit of the synthetic spectrum 
with two modified Planck curves produces
a component at 8.4 K (Fig. 11d), which is an artefact due to the combination of the
cold component at 15 K and the reference zero level emission spectrum. 
Therefore, we conclude  that, out of the Galactic plane,
 the detection of very cold dust
is due to the combination of the cold and isotropic components.
The optical depth of the very cold component found by Reach is correlated to
the optical depth of the warm component  
(Fig. 6 of Reach et al., 1995) since  
the cold molecular emission ($\sim$15 K) globally correlates 
with the warm emission (cirrus clouds).
In the Galactic plane, the detection of two correlated components 
is due to the high diversity of  physical conditions (from HII regions
to cold molecular clouds).

%----------------------------------Fig 12a
\begin{figure}  
\epsfxsize=9.cm
\epsfysize=8.cm
%\hspace{2.cm}
\vspace{0.0cm}
\epsfbox{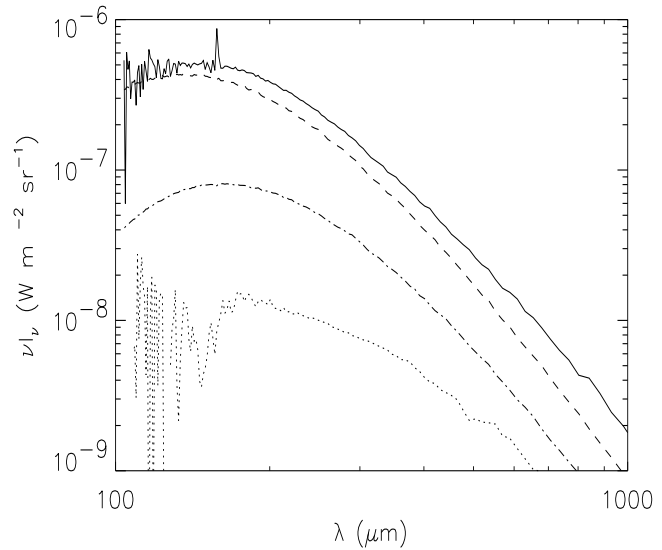}
\caption{Mean FIRAS spectrum in regions with 
significant cold emission (continuous line) is shown together with its
decomposition into the warm (dashed line), cold (dashed dotted line)
and reference zero emission (dotted line) components.}
\end{figure}

%----------------------------------Fig 12b
\begin{figure}  
\epsfxsize=9.cm
\epsfysize=8.cm
%\hspace{2.cm}
\vspace{0.0cm}
\epsfbox{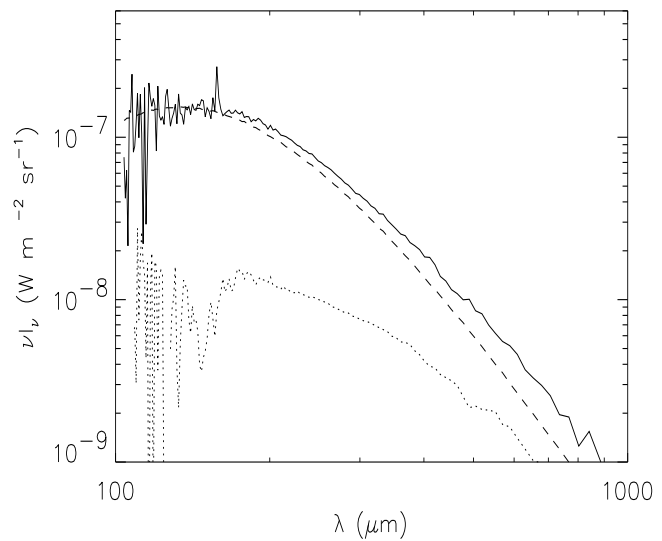}
\caption{Mean FIRAS spectrum in "cirrus" regions is shown together 
with its decomposition into the warm (dashed line) and reference zero emission
(dotted line) components.}
\end{figure}

\section{Conclusion}
We have combined the DIRBE and FIRAS data to analyse the distribution
of dust temperatures in the nearby interstellar medium seen at $|b| >10\degr$.
The main results of this work follow:\\
(1) The sky-averaged temperature of the dust in the whole Galaxy 
shows variations between lines of sight containing star forming molecular
regions, clouds forming low mass stars, and cirrus clouds.
At high latitude, the spatial variations
of the temperatures can be converted to variations in 
the radiation field intensity smaller than $\sim30\%$.\\
(2) Maps of the excess of the FIR/60$\mic$ correlation 
trace the cold component of the dust emission. 
This cold component appears to coincide with large molecular complexes 
with low star forming activity (such as Taurus,
Polaris, Chamaeleon...). The association between the cold component and molecular clouds 
is further demonstrated
by the fact that all sky pixels with significant cold emission have an 
excess IR emission with respect to the high latitude IR/HI correlation.\\
(3) The combination of DIRBE maps of the cold 
emission with FIRAS spectra have pointed out the 
significant difference in the dust temperature for the diffuse and dense parts of the 
interstellar medium
(T distributed around 17.5 K and 15 K respectively with a $\nu^2$ emissivity law in both
cases inside a beam of 7$\degr$). \\
(4) The FIRAS spectral residues of our analysis do not indicate 
any detectable very cold dust 
(4-7 K) present in the Galaxy at the level found by Reach et al. (1995).\\
(5) We conclude that the ``detection'' of such a very cold dust in our Galaxy is 
due to the presence of an isotropic background (Puget et al., 1996), together 
with the cold molecular clouds identified from DIRBE data. 
FIRAS spectra are very well decomposed by one isotropic component
and one or two Galactic components (Fig. 12 and 13).\\

\acknowledgements
We would like to thanks F.X. D\'esert, J.P Bernard, Antony Jones and E. Dwek 
for enlightening discussions.
Particular thanks go to W.T. Reach for helpfull comments and discussions.
G.L. acknowledges the hospitality of the Goddard
Space Flight Center in Maryland.
We are grateful to the Goddard Space Flight Center team for 
introducing us to the COBE data.

\end{document}